\documentclass[12pt,preprint]{aastex}
\renewcommand\>{{\rangle}}
\newcommand\bld[1]{\mbox{\boldmath $#1$}}
\newcommand\<{{\langle}}

\newcommand{\pdv}[2]{\frac{\partial#1}{\partial#2}}
\newcommand{\bnabla}{\bld{\nabla}}
\newcommand{\bB}{\bld{B}}

\newcommand{\by}{\bld{y}}
\newcommand{\bz}{\bld{z}}
\renewcommand{\bv}{\bld {v}}
\newcommand{\bk}{\bld {k}}
\newcommand{\bO}{\bld{\Omega}}
\newcommand{\uv}[1]{\hat{\bld{#1}}}
\newcommand{\vorb}{\bv_{orb}}

\newcommand{\zeus}{{\tt ZEUS }}
\newcommand{\zeusp}{{\tt ZEUS}}

\newcommand{\del}{\partial}

\newcommand{\pr}{{\rm\, Pr}}
\newcommand{\prm}{{\rm\, Pr_{M, T}}}
\newcommand{\nut}{{\rm\, \nu_T}}
\newcommand{\etat}{{\rm\, \eta_T}}

\newcommand{\be}{\begin{equation}}
\newcommand{\ee}{\end{equation}}
\newcommand{\bea}{\begin{eqnarray}}
\newcommand{\eea}{\end{eqnarray}}

\shortauthors{Guan and Gammie}
\shorttitle{}
\begin{document}

\title{The Turbulent Magnetic Prandtl Number of MHD Turbulence in Disks}

\author{Xiaoyue Guan and Charles F. Gammie\altaffilmark{1}}
\affil{Astronomy Department, University of Illinois,
1002 West Green St., Urbana, IL 61801, USA}

\altaffiltext{1}{Physics Department, University of Illinois}

\begin{abstract}

The magnetic Prandtl number ${\rm Pr_M}$ is the ratio of viscosity to
resistivity.  In astrophysical disks the diffusion of angular momentum
(viscosity) and magnetic fields (resistivity) are controlled by
turbulence.  Phenomenological models of the evolution of large scale
poloidal magnetic fields in disks suggest that the turbulent magnetic
Prandtl number $\prm$ controls the rate of escape of vertical field from
the disk; for $\prm \leq R/H$ vertical field diffuses outward before it
can be advected inward by accretion.  Here we measure field diffusion
and angular momentum transport due to MHD turbulence in a shearing box,
and thus $\prm$, by studying the evolution of a sinusoidal perturbation
in the magnetic field that is injected into a turbulent background.  We
show that the perturbation is always stable, decays approximately
exponentially, has decay rate $\propto k^2$, and that the implied $\prm
\sim 1$.

\end{abstract}


\section{Introduction}

Astrophysical disk evolution may be controlled in part by magnetic
fields that are coherent over scales of order the radius $R$.  Large
scale fields are an essential element of theoretical models for the
launching and collimation of disk winds (e.g. \citealt{bp82}; for a
recent review see \citealt{pudritz07}), which lead to disk evolution
because they exert direct torques on the surface of the disk.  Large
scale fields may also control the strength of turbulent angular momentum
diffusion (in dimensionless form, $\alpha$) within the disk, since
numerical experiments in unstratified ``shearing boxes'' suggest that
$\alpha$ is proportional to the mean field strength (e.g.
\citealt{hgb95}).  So what determines the strength of the large scale
field in well ionized disks?

A few of the physical processes affecting the large scale field strength
in thin disks can be easily listed: (1) advection of flux by large scale
flows in the disk (e.g., inward advection by accretion, but also
possibly meridional circulation); (2) turbulent diffusion; (3) dynamical
interaction between large scale fields and large scale flows (e.g.
enhanced accretion due to external torques); (4) control of small scale
MHD turbulence by large scale fields (e.g.  through the influence of a
mean field on $\alpha$); (5) generation of large scale fields by small
scale MHD turbulence (a large scale dynamo); (6) introduction or removal
of magnetic flux at the disk boundaries.  These processes are difficult
to study numerically because they involve nonsteady flows and a large
dynamic range in length scale ($R$ : scale height $H$) and time scale
(viscous timescale : dynamical timescale).

A starting point for understanding large scale field evolution is the
phenomenological model of Van Ballegooijen (1989; hereafter VB89).  He
considers a passive large scale field advected inward by accretion and
diffused by turbulence in the disk (processes [1] and [2] above).
Turbulence is modeled using a turbulent viscosity $\nut$ and turbulent
resistivity $\etat$.  The evolution of the poloidal field is then
governed by the induction equation in vector potential form:
\begin{equation}
\del_t A_\phi =
         \etat\del_R [{{1}\over{R}} \del_R (R A_\phi)]  
        + \etat \del_z^2 A_\phi                 
        + v_R {{1}\over{R}} \del_R (R A_\phi) ,
\label{INDEQ}     
\end{equation}
where $A_\phi$ is the azimuthal component of the vector potential.
$A_\phi$ labels field lines; when $\del_t A_\phi \ne 0$ the field lines
move radially through the disk with speed $\sim \del_t A_\phi/\del_R
A_\phi$.  In the limit that $\etat \rightarrow 0$ a vertical field would
be advected inward by accretion and vertical field strength would
increase with time.  In the limit that $\nu \rightarrow 0$ the field
lines simply diffuse out of the disk.  Where do advection and diffusion
balance?

VB89 give a surprising answer that can be understood as follows,
assuming the field and disk are symmetric about the midplane.  At the
midplane the first term in (\ref{INDEQ}) (radial diffusion of vertical
field) is $\sim \etat B_z/R$, where $B_z = (1/R) \del_R (R A_\phi)$ and
we assume that $\del_R \sim 1/R$.  The second term (vertical diffusion
of radial field, which can nevertheless cause radial motion of field
lines) depends on the field geometry above and below the disk.  If we
assume the field lines enter and exit the disk at an angle of order
unity (as in the wind model of \citealt{bp82}), $B_R = -\del_z A_\phi \sim
\mp B_z$ at $z = \pm H$ and $B_R = 0$ at the midplane, by symmetry. Then
$\etat \del_z^2 A_\phi \sim \etat B_z/H$. Using the usual viscous disk
estimate $v_R \sim \nu/R$, the final term (field advection) is of order
$\nu B_z/R$.  To summarize, the terms on the right hand side of
(\ref{INDEQ}) are in the ratio $\etat B_z/R : \etat B_z/H : \nu B_z/R$.
Evidently the first term is negligible in comparison to the second for a
thin disk, provided the turbulent diffusion can be described by a scalar
diffusion coefficient.  The second and third third can balance when
$\nut/\etat \sim R/H$; diffusion and advection balance when $\prm \sim
R/H$.  

It is plausible that in disk turbulence $\prm \sim 1$ (e.g.
\cite{ybr03}).  Then outward field diffusion occurs on a timescale $R
H/\nu$.  Large scale poloidal fields would vanish from the disk absent a
dynamo that regenerates the field on the same timescale (process [5]
above).  Similar conclusions have been reached by \cite{lpp94} in the
context of magnetically generated outflow/jet models.  More complex
models by \cite{hpb96} also support such a picture. 

There are of course ways to avoid the loss of large-scale field implied
by the VB89 model, which is based on a purely phenomenological model for
evolution of the disk and field.  \cite{uzspruit} discuss a model that
reduces turbulent diffusion by grouping large scale vertical magnetic
fields into bundles in the disk through flux expulsion. In these bundles
the fields are strong enough to quench turbulence and thus avoid outward
diffusion (processes [3] and [4] above).  It has also been suggested by
\cite{uzgoo} that disk atmospheres might develop loop-like large scale
coronal structures that delocalize disk evolution by transmitting
angular momentum and energy.  \cite{rl08} consider the possibility that
field diffusion is suppressed by a coronal layer where $v_A > c_s$, so
that the MRI is suppressed and $\etat$ is reduced.  Another possibility
is that disk evolution is driven by external torques associated with an
MHD (Blandford-Payne type) wind running along the poloidal field lines.
If the field is strong enough then the inflow speed $v_R$ may be large
enough to compete with outward diffusion of field lines even when $\prm
\sim 1$ (process [3] above).  \footnote{Although a naive estimate suggests
this will not work.  Suppose that the external torque per unit area
$\tau \sim R M_{z\phi} \sim (\Sigma/H) R \<v_A\>^2$, where $M_{ij}
\equiv$ magnetic stress tensor, $\<v_A\> \equiv$ the midplane
Alfv\'en speed associated with the large scale field and $\Sigma \equiv$
surface density.  Then the induced $v_r \sim \tau/(\Omega \Sigma R) \sim
\<v_A\>^2/c_s$, so in eq.(1) the second and third terms balance for
$\etat \sim \<v_A\>^2/\Omega$. If $\etat \sim \<\delta v_A^2\>/\Omega$
($\delta v_A \equiv$ the Alfv\'en speed associated with disk
turbulence), then (diffusion/advection) $\sim \<\delta
v_A^2\>/\<v_A\>^2$.  Shearing box experiments suggest that this is large
compared to $1$ unless the poloidal field is so strong that turbulence
is suppressed.}  A final possibility is that $\prm \gg 1$.

In this paper we measure the turbulent magnetic Prandtl number $\prm$
directly from shearing box simulations.  We infer the turbulent
viscosity $\nut$ from the turbulent shear stress $w_{xy,T}$ (in
dimensionless form, $\alpha$), which controls diffusive radial transport
of angular momentum in disks.  We infer the turbulent resistivity
$\etat$ by tracking the evolution of a sinusoidal disturbance in the
magnetic field that is imposed on an already turbulent state.

It is worth emphasizing that the turbulent Prandtl number $\prm$ is
fundamentally different from the Prandtl number $\pr_M$ associated with
microscopic processes (e.g., \citealt{bh08}).  Recent numerical experiments
\citep{fp07,ll07} suggest that $\pr_M$ can influence the saturation
level of MRI driven disk turbulence at low Reynolds number. 

The paper is organized as follows. In \S 2 we give a simple description
of the local model and summarize our numerical algorithm. In \S 3  we
describe the numerical procedure for measuring $\etat$.  We report
$\etat$ and discuss its dependence on the model parameters. \S 4 We
explain how we calculate $\prm$ and discuss our results.

\section{Local Model and Numerical Methods}

Our starting point is the local model for disks.  It is obtained by
expanding the equations of motion around a circular-orbiting coordinate
origin at cylindrical coordinates $(r,\phi,z) = (r_o, \Omega_o t +
\phi_o, 0)$, assuming that the peculiar velocities are comparable to the
sound speed and that the sound speed is small compared to the orbital
speed.  The local Cartesian coordinates are obtained from cylindrical
coordinates via $(x,y,z) = (r - r_o, r_o [\phi - \Omega_o t - \phi_o],
z)$.  We assume throughout that the disk is isothermal ($p = c_s^2
\rho$, where $c_s$ is constant), and that the disk orbits in a Keplerian
($1/r$) potential.

In the local model the momentum equation of ideal MHD becomes
\be\label{BE2}
\pdv{\bv}{t} + \bv\cdot \bnabla \bv + c_s^2\frac{\bnabla\rho}{\rho} + \frac{\bnabla B^2}{8\pi \rho}
- \frac{(\bB\cdot \bnabla)\bB}{4\pi \rho} + 2 \bO \times \bv - 3\Omega^2 x \, \uv{x} = 0.
\ee
The final two terms in equation (\ref{BE2}) represent the Coriolis and
tidal forces in the local frame.  Our model is unstratified, which means
that the vertical gravitational acceleration $-\Omega^2 z$ usually
present in Keplerian disks is ignored.  The box has size $L_x \times L_y
\times L_z$.

Our model contains no explicit dissipation coefficients.  Recent models
with explicit scalar dissipation (\citealt{fp07}, \citealt{ll07}) have
shown that the saturated field strength in magnetized disk turbulence
depends on the viscosity $\nu$ and resistivity $\eta$, and that \zeus
has an effective magnetic Prandtl number $\prm \equiv \nu/\eta \gtrsim
1$ (\citealt{fp07}).  

The orbital velocity in the local model is
\be
\vorb = -{3\over{2}}\Omega x \, \uv{y}.
\ee
This velocity, along with a constant density and zero magnetic field, is
a steady-state solution to equation (\ref{BE2}).  If the computational
domain extends to $|x| > (2/3) H = (2/3) c_s/\Omega$, then the orbital
speed is supersonic with respect to the grid.

The local model can be studied numerically using the ``shearing box''
boundary conditions (e.g. \citealt{hgb95}).  The boundary conditions on
the $y$ boundaries of the box are periodic, while the $x$ boundaries are
``nearly periodic'', i.e. they connect the radial boundaries in a
time-dependent way that enforces the mean shear flow.  We also use
periodic boundary conditions in the vertical direction; this is the
simplest possible version of the shearing box model.

Our models are evolved using \zeus \citep{sn92}.  \zeus is an
operator-split, finite difference scheme on a staggered mesh.  It uses
artificial viscosity to capture shocks. \footnote{A Von-Neumann
Richtmyer artificial viscosity is a pressure proportional to $(\bnabla
\cdot \bv)^2$ that is small outside shocks.  It should not be confused
with the ``anomalous viscosity'' used to model turbulent angular
momentum diffusion in standard accretion disk theories \citep{ss, lbp}.}
For the magnetic field evolution \zeus uses the Method of
Characteristics-Constrained Transport (MOC-CT) scheme, which is designed
to accurately evolve Alfv\'en waves (MOC) and to preserve the $\bnabla
\cdot \bB = 0$ constraint to machine precision (CT).

We have modified \zeus to include ``orbital advection'' \citep{mass00,
gamm01, jg05} with a magnetic field \citep{jgg08}.  Advection by the
orbital component of the velocity $\bv_{orb}$ (which may be supersonic
with respect to the grid) is done using interpolation.  With this
modification the timestep condition $\Delta t < {\it C}  \Delta
x/(|\delta \bv| + c_{max})$ ($c_{max} \equiv $ maximum wave speed and
${\it C} \equiv$ Courant number) depends only on the perturbed velocity
$\delta \bv = \bv - \bv_{orb}$ rather than $\bv$.  So when $|\bv_{orb}|
\gtrsim c_{max}$ (for shearing box models with $v_A^2/c_s^2 \lesssim 1$,
when $L \gtrsim H$) the timestep can be larger with orbital advection,
and computational efficiency is improved.

Orbital advection also improves accuracy. \zeusp, like most Eulerian
schemes, has a truncation error that increases as the speed of the fluid
increases in the grid frame.  In the shearing box without orbital
advection the truncation error would then increase monotonically with
$|x|$.  Orbital advection reduces the amplitude of the truncation error
and also makes it more nearly uniform in $|x|$ \citep{jgg08}.  

We have also implemented an additional procedure to remove the radially
dependent numerical dissipation in large shearing box simulations that
was reported in \cite{jgg08}. We do so by systematically shifting the
entire box by a few grid points in the radial direction at $t = n 2
L_y/(3 \Omega L_x$, $n = 1,2,3,\ldots$ (this is when the shearing box
boundary conditions are exactly periodic).  After the shift we execute a
divergence cleaning procedure to remove the monopoles that build up due
to truncation error along the radial boundaries of the shearing box in
our implementation of the shearing box boundary conditions, which remaps
the EMFs rather than the magnetic field.  We do this by gathering $s =
\bnabla \cdot \bB$ onto a single processor, solving the Poisson equation
$\nabla^2\psi = s$ using a standard FFT-based procedure, and then
setting $\bB \rightarrow \bB - \bnabla\psi$.  The additional
computational cost is negligible (usually less than $0.1\%$ of the total cost) because the operation is performed infrequently. \cite{jky08} have
discussed other techniques to eliminate the radial dependence of
numerical diffusion.  This procedure eliminates the features reported in
\cite{jgg08}.

\section{Turbulent Resistivity}

Our focus is on the diffusive effects of MHD turbulence induced by the
magnetorotational instability (MRI; \citealt{bh91}).  All models in this
section have a mean toroidal field $\<\bB\> = B_{0}\hat{\by}$, where
$B_{0}$ is chosen so that the initial plasma parameter $\beta \equiv
8\pi P_0/B_0^2 = 400$.  The models are evolved long enough ($\geq 40$
orbits) to reach a saturated (statistically steady) state.

First we consider radial diffusion of a vertical field.  Although this
is the subdominant term in eq.(\ref{INDEQ}), it is easier to measure
(for reasons we will discuss shortly) and thus allows us to explore
systematic effects related to resolution, scale of the perturbation, and
details of the initial state more readily.

To measure the turbulent resistivity we evolve an initial state
containing a mean field for hundreds of orbits.  We then (arbitrarily)
select an instant to inject a magnetic field perturbation of the form
\begin{equation}
\delta \bB_z = a \sin(k_x x)\hat{\bz}.
\label{delbz}
\end{equation}
Here $k_x = 2\pi n_x/L_x$ is the radial wavenumber for the perturbation.
We consider models with $4 < L_x/H < 32$.  We choose $a$ so that it is
larger than the background turbulent fluctuations 
\footnote{The background contains power in the magnetic field at small
wavenumber, with the power spectrum $\delta B_k^2 \sim const. \sim
\<\delta B^2\> \lambda^3$, where $\lambda$ is a correlation length (see
\citealt{ggsj08} for a discussion).  The implies that the Fourier
series coefficients in the background state will have rms amplitude 
$\<\delta B^2\>^{1/2} (\lambda^3/(L_x L_y L_z)^{1/2}$}, 
but small enough so that it does not greatly affect the
background turbulence.  Here we use $0.1 < a/B_0 < 0.8$.  We then evolve
the perturbed turbulence self-consistently, taking the sine transform 
of $\bB_z$ to obtain $a(t)$.

A typical evolution $a(t)$ is shown in Figure 1.  The initial decay is
approximately exponential. We measure a decay rate $1/\tau$ by fitting
an exponential to $a(t)$ over at least one e-folding time.  The decay
rates are listed in Table 1.  To give a sense of the amplitude of
fluctuations in the background state we also plot in Figure 1 the
evolution of the cosine amplitude; since the cosine amplitude remains so
much smaller than the sine amplitude it is clear that the decay of
$a(t)$ is not simply due to phase drift.  Eventually $a(t)$ decays to
the background level.  In every case we have examined the perturbation
decayed; the turbulence was stable to a large-scale magnetic field
perturbation.

Before going on we need to determine how strongly our measurement of
$\tau$ depends on our selection of initial conditions.  We therefore
measure $\tau$ for an ensemble of initial conditions selected from the
same run at widely separated times.  We have done this for three models
(s2, s5, and n1).  The variation in $\tau$ is $\leq 13\%$, which may
then be regarded as an error bar on our measurement.

The decay time $\tau$ is related to $\etat$ as follows.  Solving
the induction equation with a scalar resistivity, 
\begin{equation}
a = a_0 \exp(-\etat k_x^2 t),
\end{equation}
so we define
\begin{equation}
\etat \equiv (k_x^2 \tau)^{-1}.
\label{eta}
\end{equation}
Values of $\etat$ are given in Table 1.  In general $\etat$ will depend
on the magnitude and direction of $\bk$ and on the background field
$\<\bB\>$.

Does the decay time scale as $k_x^{-2}$, i.e. does $\etat$ depend on the
magnitude of $\bk$?  In a model with $(L_x, L_y, L_z) = (8, 4\pi, 2)H$
we imposed perturbations with $n_x = 1, 2, 3$ on the same initial state.
We find $\tau = 32.7, 10.4 , 4.93$ respectively, so that $\etat =
0.0495, 0.0388, 0.0366$,  crudely consistent with a diffusive scaling.
Models n1, s5, and s6, with $n_x = 1$ but $L_x = 8, 16, 32$ and
identical numerical resolution, have $\etat = 0.0495, 0.0330, 0.0304$,
again crudely consistent with a diffusive scaling.  So it appears that
$\etat$ is at most very weakly dependent on the magnitude of $\bk$.

We checked the effect of resolution in models with $N_x/L_x$ ranging
from $32/H$ to $128/H$.  In all the runs $\Delta x = \Delta z \simeq 2
\Delta y$ (except in model s4, where $\Delta x = \Delta z = \Delta y$;
varying the zone aspect ratio made no difference in $\etat$) and $(L_x,
L_y, L_z) = (4, 2\pi, 1)H$.  Comparing runs r1, r2, and r3, we find
$\etat = 0.037$, $0.035$ and $0.034$ respectively.  Our measurement of
$\etat$ is thus consistent with convergence, since the variation with
resolution is smaller than the noise in decay time measurement that
arise from choosing a particular initial state.

We found that $\etat$ {\em does} depend on the field perturbation
strength $a_0$. In the $(L_x, L_y, L_z)  = (8, 4\pi, 2)H$ models with
the same wavelength perturbation and a fixed resolution, when $a_0$
increases from $0.1 B_{0}$ to $0.4 B_{0}$, $\etat$ increases slightly
($\sim 20\%$) with $a_0$; when $a_0 = 0.8 B_{0}$ $\etat$ almost doubles.
This is not surprising.  It is known that the velocity fluctuation
amplitude in shearing boxes increases in the presence of a background
mean field.  For sufficiently long wavelength sinusoidal perturbations,
the imposed vertical fields looks, locally, like a mean field and so
increases the velocity fluctuation amplitude and therefore $\etat$.

We did not find any dependence of $\etat$ on box size.  Comparing runs
with $L_x/H = 4 $ and $L_y/H = 2\pi, 4\pi, 8\pi$ we found $\etat \simeq
0.03$ in every case.  Comparing runs with $L_x/H = 4$ to $32$, we again
see that $\etat$ lies in a narrow range around $0.03$. Notice that for a
vertical field perturbation with $\lambda _x < 4H$ the decay time is
less than $\Omega ^{-1}$. 

But what of the dominant term in eq(\ref{INDEQ}), the vertical diffusion
of radial field?  It is unclear how to measure this in the shearing box
because the radial field is always buried ``in the disk'', where it is
continually sheared into azimuthal field by the background shear (in a
stratified disk the radial field is present above and below the disk,
where the Alfv\'en speed is high and the plasma will corotate along
field lines).  So instead we measured the vertical diffusion of
azimuthal field.  This is also relevant to wind models because the
radial field above and below the disk is always accompanied by an
azimuthal field. 

To measure the azimuthal field diffusion we inject a perturbation of the
form
\begin{equation}
\delta \bB_y = a \sin(k_z z)\hat{\by}
\label{delby}
\end{equation}
into an already turbulent state, measure the decay time $\tau$, and set
\begin{equation}
\etat = (k_z^2 \tau)^{-1}.
\end{equation}
We imagine a large scale field entering the disk at a inclination of
$\sim 30\deg$, running vertically through the midplane, and leaving at a
similar inclination; to mimic this geometry we set $2\pi/k_z = L_z = 4
H$ and $2 H$.  

The perturbation amplitude must be chosen larger than the turbulent
background but as small as possible so it does not influence the
background state.  Because the turbulent fluctuations in the azimuthal
field are larger than the fluctuations in the vertical field, $a$ in
eq.(\ref{delby}) must be an order of magnitude larger (in comparison to
$B_0$) than $a$ in eq.(\ref{delbz}).

Model v1 and v2 (see Table 1) have $(L_x, L_y, L_z) = (4, 4\pi, 4)H$ and
resolution $128\times 200\times 128$;  v1 has $a_0 = 2B_{0}$ while v2
has $a_0 4B_{0}$.  We found $1/\tau = 0.052$ and $0.050$ respectively.
This corresponds to a vertical diffusion coefficient $\etat \sim 0.020$.
Details are given in Table 1.

To test for resolution dependence we repeated the above experiment at
resolution $256\times 400\times 256$, with $a_0 = 4B_{0}$ (model v3 in
Table 1). We found $1/\tau = 0.040$ and thus $\etat \sim 0.016$. This
diffusion coefficient is $\sim 20\%$ smaller than that obtained at lower
resolution.  This small decrease in $\etat$ is mainly caused by slightly
lower turbulent saturation level in the high resolution run in the
perturbed state, where the saturation $\alpha$ is about 25\% smaller.
This might be surprising because on average higher resolution runs have
higher saturation levels (see, e.g., Guan et al. 2009), but $\alpha$
fluctuates in time; our $\alpha$ is averaged over the same time interval
used to fit $1/\tau$.

\section{Discussion: Turbulent Magnetic Prandtl number}

To calculate $\prm$, we need to assign a ``viscosity'' to the 
turbulence.  We do this by measuring the turbulent shear stress
\begin{equation}
w_{xy,T} = \< \rho v_x \delta v_y - {B_x B_y\over{4 \pi}} \>
\end{equation}
and equating this to the shear stress that would be measured in
a viscous fluid
\begin{equation}
w_{xy,v} = \rho \nu q \Omega ,
\end{equation}
where $q \equiv -(1/2) d\ln\Omega^2/d\ln R = 3/2$ for a Keplerian
potential.  Thus
\begin{equation}
\nu = {w_{xy,T}\over{\rho q \Omega}}.
\end{equation}
and the turbulent magnetic Prandtl number
\begin{equation}
\prm \equiv {{\nut}\over{\etat}} = (\tau\Omega) (k H)^2
         {w_{xy,T} \over{q \< \rho \> c_s^2}}.
\label{prm}
\end{equation}
The turbulent shear stress $w_{xy,T}$ is related to $\alpha$ by $\alpha
\equiv w_{xy,T}/ \<\rho\>c_s^{2}$.  The evolution of $\alpha$ from one
of our runs is shown in Figure 1.

We measure the time average of $\alpha$ during the decay time, denoted
$\overline{\alpha}$ (over the same period we fit $\tau$), and use
equation (\ref{prm}) to calculate $\prm$. The results are compiled in
Table 1. The vertical field experiments give $0.35 < \prm < 0.58$. The
azimuthal field experiments give $\prm \sim 1$. 

In the current limited set of simulations we see no clear sign of $\prm$
scaling with model or numerical parameters.  This consistency is
remarkable when we look at the radial diffusion of vertical field with
different perturbation field strength $a_0$.  In the $(L_x, L_y, L_z) =
(8, 4\pi, 2)H$ models when the perturbation amplitude is strong, as in
the $a_0 = 0.8 B_0$ case, both $\eta$ and $w_{xy,T}$ double compared to
their weakly perturbed counterparts with  $a_0 = 0.1 B_0$; this doubling
is precisely what is required for a constant $\prm$.  

We have also carried out comparison experiment with slightly larger initial
toroidal field strength $B_0$ ($\beta_{0} = 100$; model b2a in Table 1).
Past numerical experiments \citep{hgb95, ggsj08} imply that the
saturation level scales linearly with $B_0$.  For this model we found
$\overline{\alpha} = 0.0568$ and $\etat = 0.0878$.  Both the turbulent
saturation level and $\etat$ double, giving $\prm \sim 0.43$. 

Recently \cite{ll08} have measured the turbulent resistivity in shearing
box simulations.  Their technique for measuring turbulent resistivity
differs from ours: they directly measure the EMF required to maintain a
particular sinusoidal variation in the field, they measure a resistive
stress tensor, and they consider only a mean vertical field (rather than
the mean azimuthal field considered here).  They find $\prm = 1.6$ for
the diffusion of a radially varying azimuthal field.  

Transport properties of the MRI-generated turbulent flow have also been
studied in the context of dust (passive scalar) mixing in a shearing box
\citep{csp05, jk05,jkm06, twby06, fp06}.  The analog of $\prm$ in these
experiments is the turbulent Schmidt number $\rm{Sc} \equiv \nu_T/D_T$,
where $D_T$ is the diffusion coefficient for the grains.  In models with
zero net magnetic flux (either when the dust is modeled as a passive
scalar \citep{twby06}, or when the dust is coupled to the gas by drag
\citep{jk05, fp06}),  $\rm{Sc} \sim 1$.  In models with a net vertical
flux \citep{csp05, jkm06}, $\rm{Sc}$ was found to be anisotropic and
increase with $\alpha$ (and therefore depend on the initial vertical
field strength),  ranging from $\rm{Sc} \sim 1$ for a weak field, to
$\rm{Sc} \sim 2$ for radial diffusion $\rm{Sc} \sim 10$ for vertical
diffusion when the mean field is strong and $\alpha \sim 0.5$.  Our
results are broadly consistent with these measurements in the sense that
${\rm Sc} \sim \prm$ when the mean field is weak.  Our scheme for measuring
$\prm$ is much more computationally expensive when the turbulence is
strong, because a large computational volume is required to reduce the
background fluctuations.  It would be interesting to investigate whether
$\prm \sim 10$ can be achieved with strong background fields in future
investigations.

A $\prm$ of order of unity is not surprising, perhaps, from a turbulent
mixing point of view.  \cite{parker79}, for example, argued that in
isotropic turbulence $\nut \sim \etat \sim l v$, where $l$ is the
largest dimension of eddies in the MHD turbulent flow and $v$ is the
characteristic eddy turnover speed. Interestingly, \cite{ybr03} also
obtained an order of unity $\prm$ from their turbulence simulations,
where the turbulence is sustained by external forcing rather than the
MRI in a disk. 

Can large scale magnetic fields avoid escape from turbulent disks?  We
clearly have not included all of the effects outlined in the
introduction that might influence the evolution of large scale fields.
Of these, perhaps the simplest route to large scale fields is a rapid
accretion mode in which the mean field torques the disk, causing it to
accrete more rapidly than would a viscous disk.  But our results cast
doubt on models that confine the large scale field by setting $\prm \sim
R/H$ (e.g. Shu et al. (2007)).

Our models do not show a clear scaling of $\prm$ with model parameters,
but the dynamic range in parameters (and thus in $\etat$) is small;
future experiments over a broader range of initial field strengths may
show scaling that is not evident here. Our models also do not include
an explicit dissipation model, on which $\prm$ might also depend.  

\acknowledgements

This work was supported by the National Science Foundation under grants
AST 00-93091, PHY 02-05155, and AST 07-09246, and a Sony Faculty
Fellowship, a University Scholar appointment, and a Richard and Margaret
Romano Professorial Scholarship to CFG. We are grateful to Shane Davis, Stu Shapiro and Fred Lamb for discussions. We also thank Ethan Vishniac and an anonymous referee for helpful suggestions. High resolution simulations were performed on
the Turing cluster at the CSE program in UIUC and Lonestar cluster at TACC.

\newpage
\begin{deluxetable}{lccccccccc}
\setlength{\tabcolsep}{0.07in}
\tabletypesize{\scriptsize}
\tablecolumns{10}
\tablecaption{Model Parameters
  \label{size}}
\tablewidth{0pt}
\tablehead{
\colhead{Model} & \colhead{Size} &
\colhead{Resolution} & \colhead{$\beta_0$} & \colhead{$a_0/B_{0}$} &
\colhead{$n_x$} &
\colhead{$\overline{\alpha}$} & \colhead{$1/\tau$} & \colhead{$\etat$} &
\colhead{$\prm$} \\
}
\startdata

\cutinhead{Radial Diffusion of a Vertical Field}

\sidehead{size}
s1   &  $(4, 2\pi, 1)H$  &  $64/H$ & 400 & 0.2   & 1
& 0.0232 
& 0.0758 & 0.0307  &  0.50\\
s2      &  $(4, 4\pi, 1)H$  &  $64/H$ & 400 & 0.2  & 1
& 0.0235 
&  0.0660& 0.0268 &  0.58\\
s3      &  $(4, 8\pi, 1)H$  &  $64/H$ & 400 & 0.2 & 1 & 0.0238
& 0.0740 & 0.0300 & 0.53 \\
s4     &$(8, 2\pi, 1)H$  &  $64/H$ & 400 & 0.2  & 1 & 0.0198
& 0.0194 & 0.0314 & 0.42  \\
s5    & $(16, 2\pi, 1) H$ & $32/H$ & 400 & 0.2 & 1
& 0.0209 & 0.00509 & 0.0330 & 0.42 \\
s6    & $(32, 2\pi, 1) H$  & $32/H$   &  400 & 0.2  & 1 &
0.0210 & 0.00117 & 0.0304 & 0.46 \\
\\
 \noalign{\vskip 1.5ex}%
 \hline
\sidehead{$n_x$}
n1   & $(8, 4\pi, 2) H$ &  $32/H$  & 400 & 0.2  & 1 &
0.0262    & 0.0306  &  0.0495  & 0.35 \\
n2  & $(8, 4\pi, 2) H$ &  $32/H$  & 400 & 0.2  & 2 &  0.0252  & 0.0957 & 0.0388 &  0.43\\
n3  & $(8, 4\pi, 2) H$ &  $32/H$  & 400 & 0.2  & 3 & 0.0240 & 0.203 & 0.0366 &0.44\\

\\
 \noalign{\vskip 1.5ex}%
 \hline
\sidehead{$a_0$}
a1  & $(8, 4\pi, 2) H$ &  $32/H$  &  400 & 0.1  & 1 & 0.0239
& 0.0268 & 0.0434  & 0.37  \\
a2   & $(8, 4\pi, 2) H$ &  $32/H$  &  400 & 0.2  & 1 &
0.0262    & 0.0306  & 0.0495  & 0.35 \\
a3  & $(8, 4\pi, 2) H$ &  $32/H$  & 400 & 0.4  & 1 & 0.0338   & 0.0315 & 0.0511  & 0.44 \\
a4   & $(8, 4\pi, 2) H$ &  $32/H$  & 400 & 0.8  & 1 & 0.0604   & 0.0496 & 0.0804  &  0.50\\

\\

 \noalign{\vskip 1.5ex}%
 \hline
\sidehead{resolution}

r1   &  $(4, 2\pi, 1)H$  &  $32/H$ & 400 & 0.4    & 1
& 0.0226 & 0.0921& 0.0373  & 0.40 \\
r2   &  $(4, 2\pi, 1)H$  &  $64/H$ & 400 & 0.4   & 1  & 0.0274
& 0.0875 & 0.0355  & 0.51 \\
r3   &  $(4, 2\pi, 1)H$  &  $128/H$ & 400 & 0.4   & 1 &  0.0239  & 0.0840 & 0.0340 & 0.47 \\

\\
 \noalign{\vskip 1.5ex}%
 \hline
\sidehead{$\beta _0$}
b1   & $(8, 4\pi, 2) H$ &  $32/H$  &  400 & 0.2  & 1 &
0.0262    & 0.0306  & 0.0495  & 0.35 \\
b2a  & $(8, 4\pi, 2) H$ &  $32/H$  &  100 & 0.2  & 1 &
0.0568     & 0.0541   & 0.0878 & 0.43 \\
b2b  & $(8, 4\pi, 2) H$ &  $32/H$  &  100 & 0.4 & 1 & 0.0607
     & 0.0463  & 0.0750 & 0.54 \\

\\
\cutinhead{Vertical Diffusion of an Azimuthal Field}

v1   &  $(4, 4\pi, 4)H$  &  $32/H$ & 400 & 2.0    & 1
& 0.0313
& 0.0521 & 0.0211  & 0.99 \\
v2   &  $(4, 4\pi, 4)H$  &  $32/H$ & 400 & 4.0   & 1  & 0.0386
& 0.0503 & 0.0204  & 1.26 \\
v3   &  $(4, 4\pi, 4)H$  &  $64/H$ & 400 & 4.0   & 1  & 0.0288
& 0.0400 & 0.0162  & 1.18 \\
v4   &  $(8, 4\pi, 2)H$  &  $32/H$ & 400 & 4.0   & 1 &  0.0276  & 0.1425 & 0.014 & 1.27 \\

\enddata
\end{deluxetable}

\newpage

\begin{figure}
\plotone{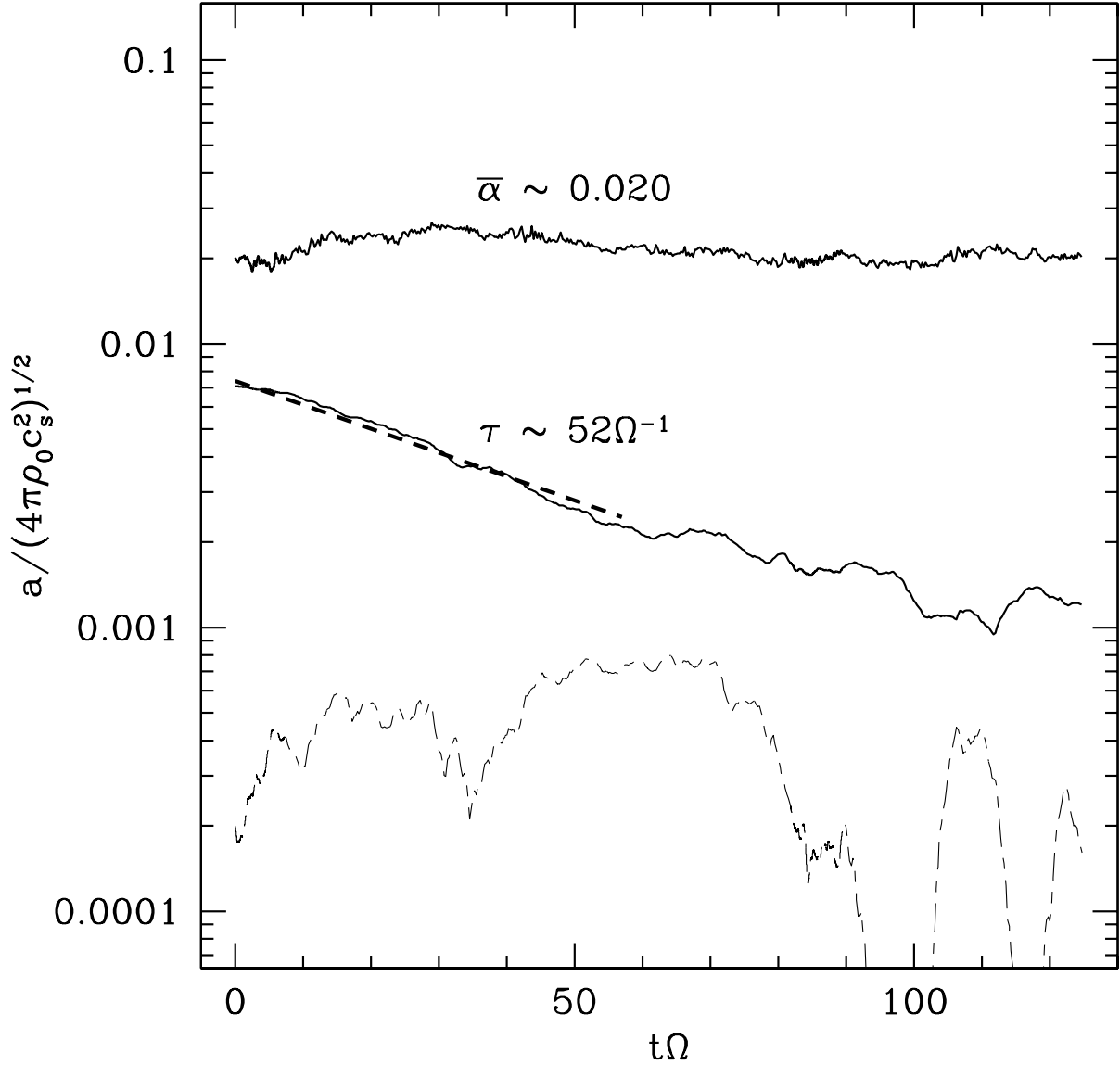}
\caption{
Evolution of the amplitude $a$ of a magnetic field perturbation $\delta
B_z = a \sin(k_x x)$ in the presence of MHD turbulence (lower solid
line; model s4).  A linear fit to $\ln a(t)$ is shown as a heavy dashed
line.  The decay time is $\tau \sim 52\Omega ^{-1}$, implying a
turbulent resistivity $\etat \sim 0.031$.  Also shown is the evolution
of $\alpha$ in the same experiment (upper solid line).  The turbulent
magnetic Prandtl number $\prm \sim 0.42$.  The dashed line shows the
evolution of the corresponding cosine amplitude in $\delta B_z$,
indicating the amplitude of large scale field fluctuations in the
turbulent background.
}
\label{fig:std.eb.vs.t}
\end{figure}

\end{document}